\documentclass[aps,prl,twocolumn,amsmath,amssymb,footinbib,showpacs,longbibliography,superscriptaddress]{revtex4-1}

\newcommand{\Jnature}{Nature (London)}
\newcommand{\Jnatphys}{Nat. Phys.}

\newcommand{\Jscience}{Science}

\newcommand{\Jprl}{Phys. Rev. Lett.}

\newcommand{\Jpra}{Phys. Rev. A}

\newcommand{\Jpre}{Phys. Rev. E}

\newcommand{\Jnjp}{New J. Phys.}

\newcommand{\JphyslettA}{J. Phys. Lett. A}

\newcommand{\Jannphys}{Ann. Phys. (NY)}

\newcommand{\JMathPhys}{J. Math. Phys.}

\usepackage{times}

\usepackage[english]{babel}
\usepackage{latexsym}
\usepackage{graphics}
\usepackage{subfigure}
\usepackage{epsfig}
\usepackage{color}
\usepackage{hyperref}
\usepackage{braket} 
\usepackage[T1]{fontenc}
\usepackage[latin9]{inputenc}
\usepackage{amstext}
\usepackage{amssymb}
\usepackage{graphicx}
\usepackage{esint}
\usepackage{braket}
\usepackage{babel}
\usepackage{amsmath}

\hypersetup{
colorlinks=true,
citecolor=blue,
linkcolor=red,
urlcolor=black
}


\newcommand{\e}{\textrm{e}}
\newcommand{\ie}{i.e.}
\newcommand{\eg}{e.g.}

\newcommand{\kB}{k_\textrm{\tiny B}}
\newcommand{\lambdadB}{\lambda_{\tiny T}}
\newcommand{\lambdaT}{\lambda_{\tiny T}}

\newcommand{\aOneD}{a_{\textrm{\tiny 1D}}}

\newcommand{\aHO}{a_{\textrm{\tiny ho}}}
\newcommand{\lHO}{a_{\textrm{\tiny ho}}}
\newcommand{\xiT}{\xi_{\textrm{\tiny T}}}
\newcommand{\xig}{\xi_{{\tiny \gamma}}}

\newcommand{\LTF}{L_{\textrm{\tiny TF}}}
\newcommand{\Lth}{L_{\textrm{\tiny th}}}

\newcommand{\lettersection}[1]{\paragraph*{#1.---}}

\begin{document}

\title{
Tan's Contact for Trapped Lieb-Liniger Bosons at Finite Temperature
}

\author{Hepeng Yao}
\affiliation{
CPHT, Ecole Polytechnique, CNRS, Universit\'e Paris-Saclay, Route de Saclay, 91128 Palaiseau, France
}

\author{David Cl\'ement}
\affiliation{
 Laboratoire Charles Fabry,
 Institut d'Optique, CNRS, Universit\'e Paris-Saclay,
 2 avenue Augustin Fresnel,
 F-91127 Palaiseau cedex, France
}

\author{Anna Minguzzi}
\affiliation{
 Univ. Grenoble-Alpes, CNRS, LPMMC, F-38000 Grenoble, France}

\author{Patrizia Vignolo}
\affiliation{
 Universit\'e C\^ote d'Azur, CNRS, Institut de Physique de Nice, 1361 route des Lucioles, 06560 Valbonne, France
}

\author{Laurent Sanchez-Palencia}
\affiliation{
CPHT, Ecole Polytechnique, CNRS, Universit\'e Paris-Saclay, Route de Saclay, 91128 Palaiseau, France
}

\date{\today}

\begin{abstract}
The universal Tan relations connect a variety of microscopic features of many-body quantum systems with two-body contact interactions to a single quantity, called the contact. The latter has become pivotal in the description of quantum gases.
We provide a complete characterization  of the Tan contact of the harmonically trapped Lieb-Liniger gas for arbitrary interactions and temperature.
Combining thermal Bethe ansatz, local-density approximation, and exact quantum Monte Carlo calculations,
we show that the contact is a universal function of only two scaling parameters, and determine the scaling function.
We find that the temperature dependence of the contact, or equivalently the interaction dependence of the entropy, displays a maximum. The presence of this maximum  provides an unequivocal signature of the crossover to the fermionized regime and it is accessible in current experiments.
\end{abstract}

\maketitle

Describing strongly correlated quantum systems from microscopic models and first principles is a central challenge for modern many-body physics. The derivation of universal relations in systems governed by contact interactions is an example of such an approach~\cite{tan2008b, tan2008c}. Pointlike interactions induce a characteristic singularity of the many-body wavefunction at short interparticle distance and, correspondingly, algebraically decaying momentum tails, $n(k) \simeq C/k^4$~\cite{olshanii2003,tan2008a}. The $1/k^4$ scaling is universal and holds irrespective of the quantum statistics, dimension, temperature, and interaction strength.
Furthermore, the weight $C$  of the tails, known as Tan's contact, contains a wealth of information about many quantities characterizing the specific state,
\eg\ the interaction energy, the pair correlation function, the free-energy dependence on interactions, and the relation between pressure and energy density~\cite{tan2008a,tan2008b,tan2008c}.
Stemming from the unique possibility to measure it in ultracold gases, the contact has become central to the description of quantum gases.
Recent experiments on three-dimensional Fermi and Bose gases have permitted us to validate the universal Tan relations, hence demonstrating that $C$ provides valuable information on a variety of thermodynamic quantities~\cite{stewart2010, wild2012, sagi2012, hoinka2013, luciuk2016, fletcher2017, laurent2017}.

Interacting one-dimensional (1D) bosons display  very different physical regimes at varying interaction strength, from quasicondensates to the emblematic \textit{fermionization} effect~\cite{girardeau1960}. So far, the emergence of statistical transmutation in the Tonks-Girardeau regime~\cite{paredes2004,kinoshita2004}, the suppression of pair correlations~\cite{kinoshita2005,jacqmin2011}, and the observation of quantum criticality~\cite{yang2017} have been reported in ultracold atom experiments.
However, the experimental characterization of the various quantum degeneracy regimes at finite temperature, identified in Ref.~\cite{petrov2000b}, remains challenging.
A major difficulty is that most quantities show a smooth monotonic behavior when crossing over different regimes.
Understanding whether the contact can provide an efficient probe is one of the motivations of our work.

In most experimental conditions, the gases are confined in longitudinal harmonic traps and thermal effects cannot be neglected.
While the homogeneous 1D gas is exactly solvable by Bethe ansatz, the trapped system is not integrable, therefore requiring approximate or {\it ab initio} numerical approaches. Previous theoretical studies have investigated the contact for homogeneous bosons at finite temperature~\cite{kheruntsyan2003,kormos2009}, trapped bosons at zero temperature~\cite{minguzzi2002,olshanii2003}, and at finite temperature in the Tonks-Girardeau limit~\cite{vignolo2013}. Momentum distributions of strongly interacting, trapped bosons at finite temperature were also computed by quantum Monte Carlo methods~\cite{xu2015}.

 \begin{figure}[b!]
\vspace{-0.5cm}
\includegraphics[width = 1.0\columnwidth,height=18em]{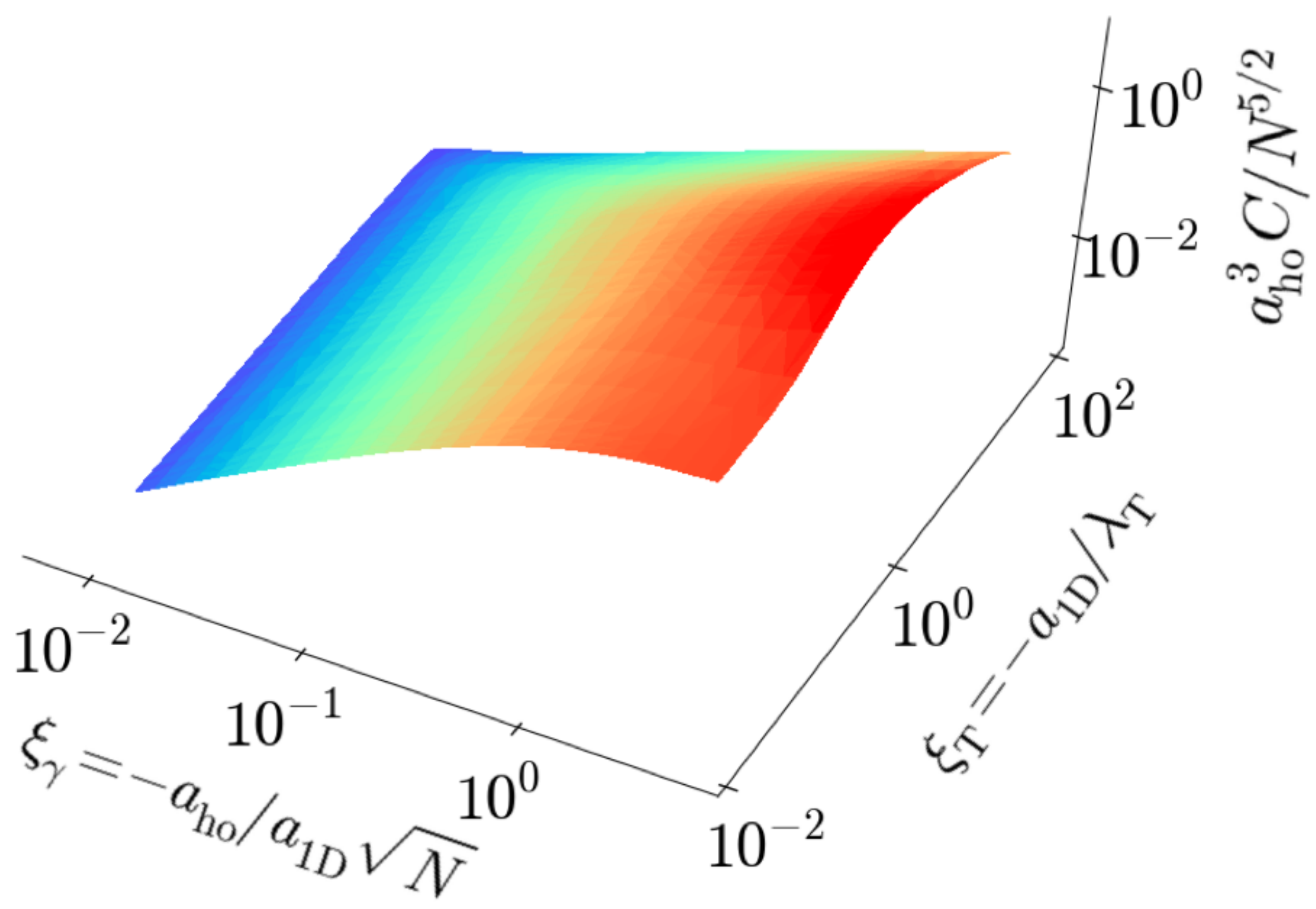}
\vspace{-0.5cm}
\caption{\label{fig:contact3D}
Reduced Tan contact $\aHO^3C/N^{5/2}$ for 1D Bose gases in a harmonic trap, versus the reduced temperature $\xiT=-\aOneD/\lambdadB$ and the reduced interaction strength $\xig=-\aHO/\aOneD\sqrt{N}$. The results are found using thermal Bethe ansatz solutions combined with local-density approximation (see main text).
}
 \end{figure}

In this Letter, we provide a complete characterization of Tan's contact of 1D bosons under harmonic confinement for arbitrary interactions, particle number, temperature, and trap frequency,
and show that it indeed provides a useful probe of quantum degeneracy regimes.
Using a combination of thermal Bethe-ansatz solutions with local-density approximation and exact quantum Monte Carlo calculations,
we demonstrate that the contact is a universal function of only two scaling parameters and find the scaling function (see Fig.~\ref{fig:contact3D}),
hence generalizing the results of Ref.~\cite{xu2015}.
As a main result, we find that the contact displays a maximum versus the temperature.
This behavior is characteristic of the trapped gas with finite, although possibly arbitrarily strong, interactions.
We argue that the existence of this maximum is a direct consequence of the dramatic change of correlations
and thus provides an unequivocal signature of the crossover to fermionization in the trapped 1D Bose gas.
We derive asymptotic limits and discuss a physical picture of the evolution of the contact versus temperature and interaction strength.
Finally,  we compute the full momentum distributions in various regimes. They show the emergence of the high-momentum tails and assess the experimental observability of our predictions.

\lettersection{Two-parameter scaling}
Consider a 1D Bose gas with repulsive two-body contact interactions, in the presence of the harmonic potential $V(x)=m\omega^2 x^2/2$, with $m$ the particle mass, $x$ the space coordinate, and $\omega/2\pi$ the trap frequency.
It is governed by the extended Lieb-Liniger (LL) Hamiltonian
\begin{equation}\label{eq:Hamiltonian}
\mathcal{H} =\sum_{j} \Big[-\frac{\hbar^2}{2m}\frac{\partial^2}{\partial x_j^2}+V(x_j)\Big]+g\sum_{j<\ell}\delta(x_j-x_\ell),
\end{equation}
where $j$ and $\ell$ span the set of particles,
and $g=-2\hbar^2/m\aOneD$ is the coupling constant with $\aOneD$ the 1D scattering length.
The thermodynamic properties of the interacting gas at the finite temperature $T$ are uniquely determined by the grand potential
$\Omega=-\kB T
\ln \left[\textrm{Tr}\,\,
\e^{ - (\mathcal{H} - \mu \mathcal{N})/\kB T} \right]$,
where $\kB$ is the Boltzmann constant,
$\mathcal{N}$ the particle number operator,
and $\mu$ the chemical potential.

We start from the homogeneous case, $V(x)=0$. Using $\kB T$ as the unit energy and, correspondingly, the thermal de Broglie wavelength $\lambdadB=\sqrt{2 \pi \hbar^2/m\kB T}$ as the unit length, we readily find that the Hamiltonian $\mathcal{H}/\kB T$ is a function of the unique parameter $\aOneD/\lambdadB$. Since the interactions are short range,
$\Omega$ is an extensive quantity. It follows that the dimensionless quantity $\Omega/\kB T$ is a function of the sole intensive, dimensionless parameters $\mu/\kB T$ and $\aOneD/\lambdadB$, times the length ratio $L/\lambdadB$. We may thus write
\begin{equation}\label{eq:GrandPotHomo}
{\Omega}/{\kB T}= ({L}/{\lambdadB})\, \mathcal{A}_\textrm{h} \big( {\mu}/{\kB T}, {\aOneD}/{\lambdadB} \big),
\end{equation}
with $\mathcal{A}_h$ a dimensionless function.

For the gas under harmonic confinement, the additional energy scale $\hbar \omega$ emerges, associated with the length scale $\aHO=\sqrt{\hbar/m\omega}$.
Within the local-density approximation (LDA), we write the grand potential
as the sum of the contributions of slices
of homogeneous LL gases with a chemical potential locally shifted by the trap potential energy,
$\Omega/\kB T = \int \frac{d x}{\lambdadB}\, \mathcal{A}_h \left[\mu-V(x),T,g\right]$. Using
Eq.~(\ref{eq:GrandPotHomo}) and rescaling the position $x$ by the quantity $2\sqrt{\pi} \aHO^2/\lambdadB$, we then find
\begin{equation}
\label{eq:GrandPotTrap}
{\Omega}/{\kB T} = \big({\aHO}/{\lambdadB}\big)^{2} \mathcal{A} \big( {\mu}/{\kB T}, {\aOneD}/{\lambdadB}  \big),
\end{equation}
with $\mathcal{A}$ a dimensionless function stemming from $\mathcal{A}_h$.
The scaling forms of the relevant thermodynamic quantities are then readily found from Eq.~(\ref{eq:GrandPotTrap}).
On the one hand, the average particle number $N=-\left.{\partial\Omega}/{\partial \mu}\right\vert_{T,\aOneD}$ reads 
\begin{equation}\label{eq:NTrap}
N = ({\aHO}/{\lambdadB})^{2} \mathcal{A}_N \big( {\mu}/{\kB T}, {\aOneD}/{\lambdadB} \big).
\end{equation}
It follows that the reduced chemical potential $\mu/\kB T$ is a universal function of only two scaling parameters,
namely
$N\lambdadB^{2}/\aHO^{2}$ and $\aOneD/\lambdadB$,
or, equivalently, $\xig=-\aHO/\aOneD\sqrt{N}$, and $\xiT=-\aOneD/\lambdadB$.
On the other hand, the contact is expressed using the Tan sweep relation~\cite{tan2008a,barth2011,note:normC},
$C=({4m}/{\hbar^2}) \left.{\partial\Omega}/{\partial \aOneD}\right\vert_{T,\mu}$, yielding
$C = ({\aHO^{2}}/{\aOneD^{5}}) \mathcal{A}_C \big( {\mu}/{\kB T}, {\aOneD}/{\lambdadB})$.
Using Eq.~(\ref{eq:NTrap}) and writing $\mu/\kB T$ as a function of $\xig$ and $\xiT$,
we find
\begin{equation}\label{eq:scaling}
C = \frac{N^{5/2}}{\aHO^{3}}  f \left(\xig, \xiT \right),
\end{equation}
with $f$ a dimensionless function.
In the following, we shall use this two-parameter scaling form.
Note that the procedure used to find the scaling form~(\ref{eq:scaling}) is general and can be straightforwardly extended to higher dimensions and Fermi gases.

\lettersection{Scaling function for 1D bosons at finite temperature}
In order to verify the scaling form~(\ref{eq:scaling}) and find the scaling function $f$ for interacting 1D bosons, we use two complementary approaches. 

On the one hand, we perform the LDA on the exact solutions of the Yang-Yang (YY) equations~\cite{yang1969}, found by the thermal Bethe ansatz for the grand-potential density
\begin{equation}
\label{omega-hom}
\Omega/L = -\kB T \int \frac{dq}{2 \pi} \ln \left[1+ e^{-\frac{\epsilon(q)}{\kB T}}\right],
\end{equation} 
and the dressed energy,
\begin{equation}
\label{epsilon-hom}
\epsilon(k)= \frac{\hbar^2 k^2}{2m} - \mu -\kB T\int \frac{dq}{2 \pi} \frac{2c}{c^2+(k-q)^2} \ln\left[1+ e^{-\frac{\epsilon(q)}{\kB T}}\right],
\end{equation}  
with $c=m g/\hbar^2=-2/\aOneD$.
We thus find the grand potential in the harmonic trap, and the scaling function $\mathcal{A}$ in Eq.~(\ref{eq:GrandPotTrap}).
Applying the procedure presented above, we then find the scaling function $f$ in Eq.~(\ref{eq:scaling}) for the contact. 

On the other hand, to assess the accuracy of the LDA, we perform \textit{ab initio} quantum Monte Carlo (QMC) calculations. We use the same implementation as in Refs.~\cite{carleo2013,boeris2016}, which allows for numerically exact simulation of the Hamiltonian~(\ref{eq:Hamiltonian}) within the grand-canonical ensemble. It yields the total number of particles $N$, the interaction energy $\langle \mathcal{H}_{\mathrm{int}}\rangle$, and the contact via the thermodynamic relation
$C= ({2gm^2}/{\hbar^4})\langle \mathcal{H}_{\mathrm{int}}\rangle$
versus temperature and chemical potential~\cite{note:SM}.

The scaling function $f$, namely the rescaled contact $\lHO^3C/N^{5/2}$, for 1D bosons under harmonic confinement resulting from YY theory and LDA is shown in Fig.~\ref{fig:contact3D}.
 \begin{figure}[t!]
 \includegraphics[width = 1.00\columnwidth]{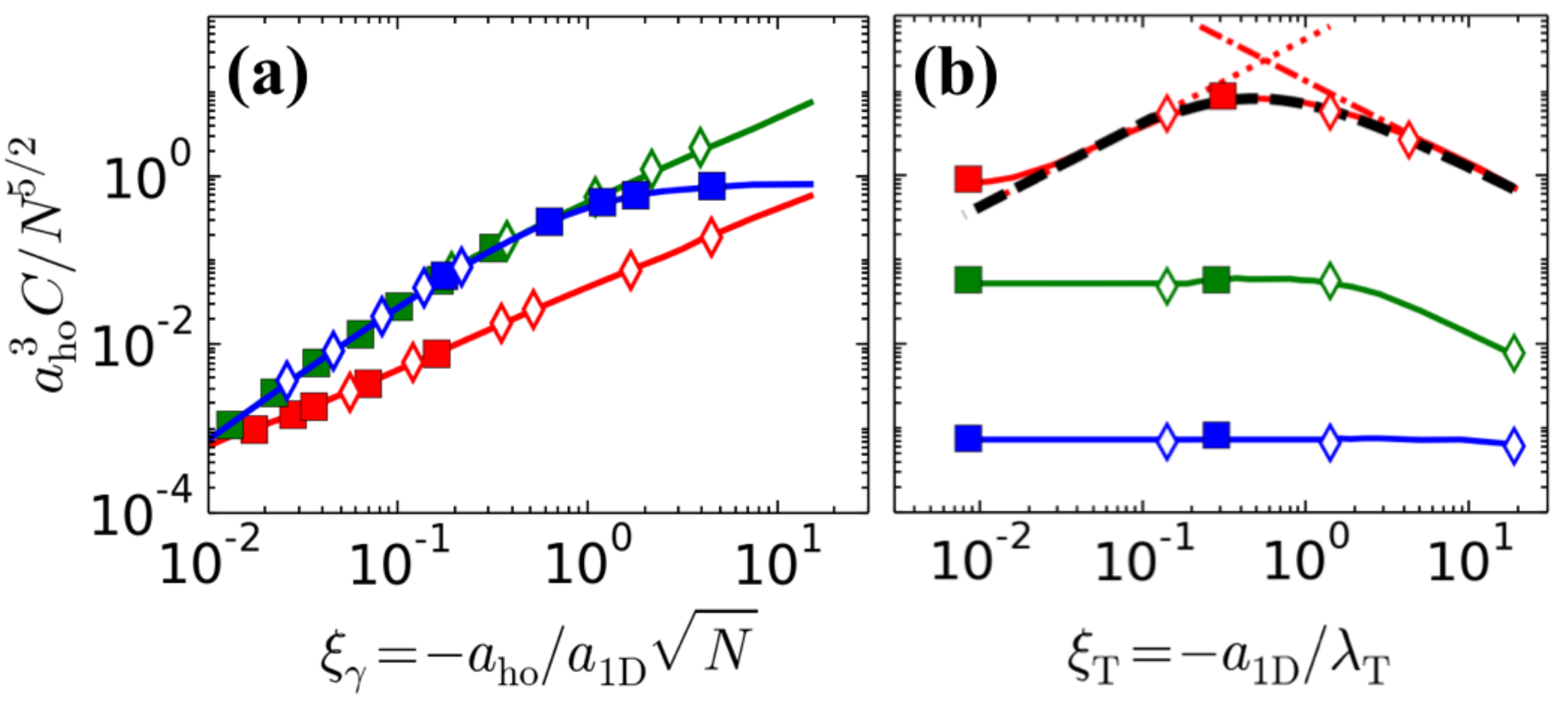}
 \caption{\label{fig:contact_vs_N-T}
Reduced Tan contact $\aHO^3C/N^{5/2}$ versus the scaling parameters,
as found from LDA (solid lines) and QMC calculations (points).
(a)~Reduced contact versus the interaction, $\xig=-\aHO/\aOneD\sqrt{N}$, at the fixed temperatures $\xiT=-\aOneD/\lambdadB = 0.0085$ (blue), $0.28$ (green), and $18.8$ (red).
(b)~Reduced contact versus the temperature via the quantity $\xiT$ at the fixed interaction strengths
$\xig=10^{-2}$ (blue), $1.58\times10^{-1}$ (green), and $15.0$ (red). The black dashed, red dotted, and red dash-dotted lines correspond to Eqs. (\ref{formula-pat}),(\ref{eq:FermionizeHighT}), and (\ref{eq:idealBoltzmans}) respectively.
The QMC data are found from various sets of parameters, corresponding to the various symbols~\cite{note:symbols}.
}
 \end{figure}
Figure~\ref{fig:contact_vs_N-T} shows some sections of the latter (solid lines) along with QMC data (points) for a quantitative comparison.
The rescaled contact is plotted as a function of the interaction strength $\xig$ for various values of the temperature via the quantity $\xiT$ in Fig.~\ref{fig:contact_vs_N-T}(a) and, inversely,
as a function of $\xiT$ for various values of $\xig$
in Fig.~\ref{fig:contact_vs_N-T}(b).
The numerically exact QMC data are computed for a broad set of parameters.
When plotted in the rescaled units of Eq.~(\ref{eq:scaling}), they show excellent data collapse among each other and fall onto the LDA curves. Quite remarkably, the agreement holds
also in the low-temperature and strongly interacting regime where the particle number is as small as $N \simeq 5$, within less that $3\%$.
Our analysis hence validates the scaling form~(\ref{eq:scaling}) and shows that the LDA is very accurate in computing the contact for the trapped LL model.

\begin{figure}[t!]
\includegraphics[width = 0.7\columnwidth]{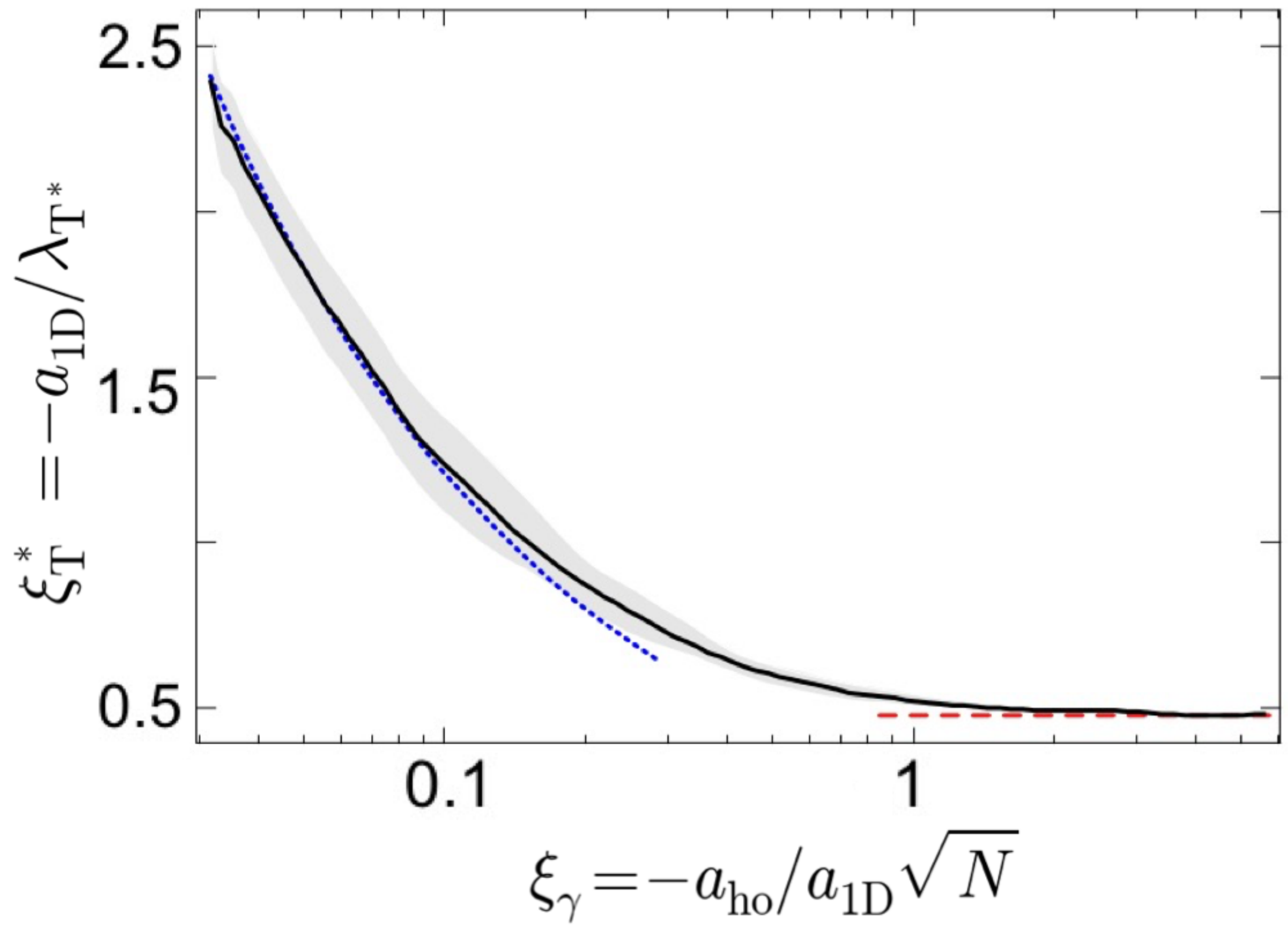}
\caption{\label{fig:maximum}
Behavior of the temperature at which the contact is maximum versus the interaction strength.
Shown is the value of $\xiT^*$ (solid black line with shaded gray error bars) as found from the data of Fig.~\ref{fig:contact3D},
together with the asymptotic behaviors $\xiT^* \simeq 0.49$ for the strongly interacting regime (dashed red line)
and $\xiT^* \propto \xiT^{\nu_\textrm{\tiny fit}}$, with $\nu_\textrm{\tiny fit} \simeq 0.6$ for the weakly interacting regime (dotted blue line). }
\end{figure}

\lettersection{Onset of a maximum contact versus temperature}
We now turn to the behavior of the contact.
Particularly interesting is the nonmonotonicity of $C$ versus temperature and the onset of a maximum, see Fig.~\ref{fig:contact_vs_N-T}(b).
This behavior strongly contrasts with that found for the homogeneous gas and the trapped gas in the Tonks-Girardeau limit ($\aOneD \rightarrow 0$), which are both characterized by a systematic increase of the contact versus temperature~\cite{kheruntsyan2003,vignolo2013}.
In the trapped case, the maximum in the contact as a function of $\xiT$ is found irrespective to the strength of interactions but is significantly more pronounced in the strongly interacting regime.
From the data of Fig.~\ref{fig:contact3D}, we extract the temperature $T^*$ at which the contact is maximum at fixed $\xig$.  In Fig.~\ref{fig:maximum}, we plot $\xiT^*=-\aOneD/\lambda_{T}^*$ as a function of $\xig$. As we discuss now, $\xiT^*$ shows significantly different behavior in the strongly and weakly interacting regimes, but, in both cases, it characterizes the onset of the regime dominated by interactions.

We first consider the strongly interacting regime, $\rho(0)|\aOneD| \lesssim 1$.
Using the virial expansion, we obtain the analytical expression for the contact~\cite{note:SM}
 \begin{equation}
    C=\dfrac{2N^{5/2}}{\pi\lHO^3}\dfrac{\xig}{\xiT}\left(\sqrt{2}-\dfrac{\mathrm{e}^{1/2\pi\xiT^2}}{\xiT}{\rm Erfc}(1/\sqrt{2\pi}\xiT)\right),
    \label{formula-pat}
 \end{equation}
see black dashed line in Fig. \ref{fig:contact_vs_N-T}(b). It has a maximum at $\xiT^*=0.485$, in very good agreement with the asymptotic scaling $\xiT^* \simeq 0.490 \pm 0.005$ extracted from the data (dashed red line in Fig.~\ref{fig:maximum}).

The existence of a maximum of the contact in the strongly interacting regime can be inferred from the competition of two different behaviors. On the one hand, at low temperatures, $|\aOneD| \lesssim \lambdadB$ ($\xiT \lesssim 1$), both quantum and thermal fluctuations are dominated by repulsive interactions and the gas is fermionized. The contact is then found from the Bose-Fermi mapping, \ie\ $C=({2\hbar^2}/{gm})\int dx\, \rho(x) e_{\tiny K}(x)$, where $e_{\tiny K}$ is the kinetic-energy density~\cite{cazalilla2003one}. Since the gas is  weakly degenerate, the latter follows from the equipartition theorem,
\ie\ $e_\textrm{\tiny K}(x)=\rho(x)\kB T /2$, and the density profile can be taken as noninteracting, \ie\
$\rho(x)=({N}/{\sqrt{2\pi}\Lth})\exp(-x^2/2\Lth^2)$, with  $\Lth=\sqrt{\kB T/m\omega^2}$.
It yields
\begin{equation}\label{eq:FermionizeHighT}
C={2\sqrt{2} N^{5/2}\xig\xiT}/{\lHO^3}, \quad \xig^{-1} \lesssim \xiT \lesssim 1,
\end{equation}
thus recovering the results of  Ref.~\cite{vignolo2013} by a different approach.

On the other hand, at high temperature, $\lambdadB \lesssim |\aOneD|$ ($\xiT \gtrsim 1$),  the weakly degenerate Bose  gas is dominated by thermal fluctuations. In this case, the contact can be estimated by the mean-field expression $\langle \mathcal{H}_{\mathrm{int}}\rangle = g \int dx\, [\rho(x)]^2$.
Using the thermal density profile, we then find
\begin{equation}\label{eq:idealBoltzmans}
C \simeq {2\sqrt{2}N^{5/2}\xig}/{\pi\xiT\lHO^3}, \quad \xig^{-1}, 1 \lesssim \sqrt{\xiT}.
\end{equation}
Both Eqs.~(\ref{eq:FermionizeHighT}) and (\ref{eq:idealBoltzmans}) are in good agreement with the numerical calculations, see red-dotted and dash-dotted lines in  Fig.~\ref{fig:contact_vs_N-T}(b).
These expressions show that the contact increases with  temperature in the fermionized regime but decreases when thermal fluctuations dominate over interactions. The maximum of the contact thus provides a nonambiguous signature of the crossover to fermionization.

The situation is completely different in the weakly interacting regime, $\rho(0) |\aOneD| \gtrsim 1$.
In this case, the gas is  never fermionized.  At low temperature, $(|\aOneD|/\rho(0)^3)^{1/4} \lesssim \lambdadB$, the gas forms a quasicondensate characterized by suppressed density fluctuations~\cite{kheruntsyan2003}.
The contact is then found from the mean-field expression for $\langle \mathcal{H}_\textrm{\tiny int}\rangle$, using the Thomas-Fermi (TF) density profile
$\rho(x) = ({\mu}/{g})(1-x^2/\LTF^2)$ with $\LTF=\sqrt{2\mu/m\omega^2}$. In this regime  one has~\cite{olshanii2003}
\begin{equation}\label{eq:quasiBEC}
C = \eta N^{5/2}\xig^{5/3}/\lHO^{3}, \quad 1, \xiT \lesssim \xig^{-1}
\end{equation}
with $\eta = 4\times 3^{2/3}/5$.

In the high-temperature and weakly interacting regime, $\lambdadB \lesssim (|\aOneD|/\rho(0)^3)^{1/4}$,
the interactions are negligible and the bosons form a nearly ideal degenerate gas.
Using the corresponding density profile
$\rho(x) = \lambdadB^{-1}\textrm{Li}_{1/2}\left[\exp\left(\alpha-x^2/2\Lth^2\right)\right]$
with $\alpha (\xig,\xiT) = \ln [1-\exp(-1/(2\pi\xig^2\xiT^2 ))]$, we find
\begin{equation}\label{eq:idealBose}
C= \left({16\sqrt{\pi}N^{5/2}\xig^5\xiT^3}/{\lHO^3}\right)\, G (\alpha),
\quad \xig^{-1} \lesssim \xiT \lesssim \xig^{-2}
\end{equation}
with $G (\alpha) = \int dx\ \textrm{Li}_{1/2}^2[\exp(\alpha-x^2)]$.
The function $G(\alpha)$ decays at least as $\lambdadB^4$ and thus $C$ decreases with the temperature.
Therefore, in the weakly-interacting regime, the maximum contact signals the crossover from the quasicondensate regime to the ideal Bose gas regime.
The position of the maximum of the contact  may be estimated by equating Eqs.~(\ref{eq:quasiBEC}) and (\ref{eq:idealBose}).
The calculation is significantly simplified by neglecting quantum degeneracy effects in Eq.~(\ref{eq:idealBose}). Then, $G(\alpha) \simeq \sqrt{\pi/2}\exp(2\alpha)$ and we find
\begin{equation}\label{eq:tweak}
\xiT^* \sim \xig^{-\nu}, \quad \nu = 2/3.
\end{equation}
The numerical data are well fitted by Eq.~(\ref{eq:tweak}) with $\nu$ as an adjustable parameter (see dotted blue line in Fig.~\ref{fig:maximum}), yielding $\nu_\textrm{\tiny fit} = 0.6 \pm 0.06$, in good agreement with the theoretical estimate $\nu_\textrm{\tiny th} = 2/3$~\cite{note:maximum-QD}.

\lettersection{Maximum entropy versus interaction strength}
To further interpret the onset of a maximum contact versus temperature, we note that it is equivalent to the onset of a maximum entropy $S$ versus interaction strength. For fixed number of particles, it is a direct consequence of the Maxwell identity~\cite{note:Maxwell}
\begin{equation}\label{eq:Maxwell}
\left.{\partial C}/{\partial T}\right\vert_{\aOneD,N} = (4m/\hbar^2) \left.{\partial S}/{\partial \aOneD}\right\vert_{T,N}.
\end{equation}
In the homogeneous LL gas, the entropy at fixed temperature and number of particles decreases monotonically versus the interaction strength, since repulsive interactions inhibit
the overlap between the particle wavefunctions, hence diminishing the number of available configurations.
In the trapped gas, however, this effect competes with the interaction dependence of the available volume.
More precisely, starting from the noninteracting regime, the system size increases sharply with  interaction strength, while the particle overlap varies smoothly. In this regime, the number of available configurations and the entropy thus increase with the interaction strength.
At the onset of fermionization, interaction-induced spatial exclusion becomes dramatic and the particles strongly avoid each other. In turn, as opposed to the noninteracting regime, the volume increases very slightly.
In this regime, the number of available configurations thus decreases when the interactions increase.
This picture confirms that the maximum of the entropy as a function of the interaction strength, or equivalently the
maximum of the contact as a function of the temperature, signals the fermionization crossover.

\lettersection{Experimental observability}

 \begin{figure}[t]
 \includegraphics[width = 1.00\columnwidth]{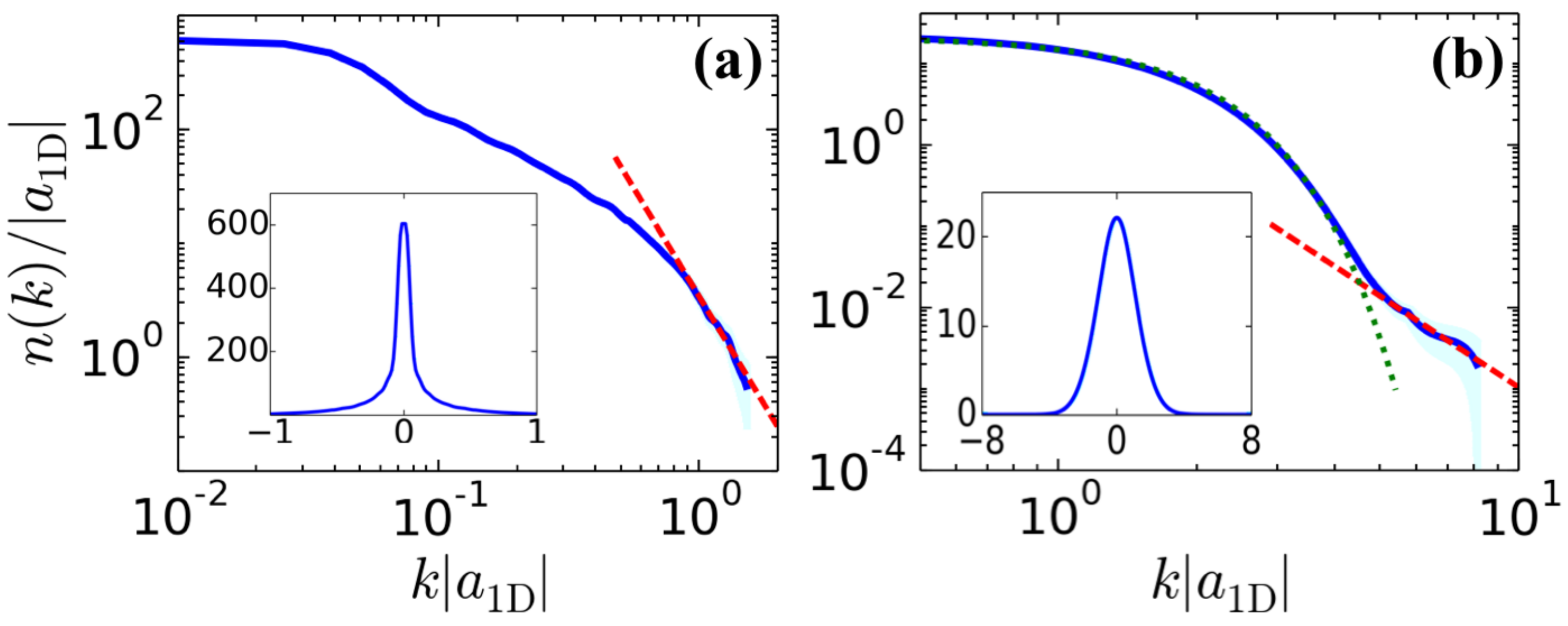}
 \caption{\label{fig:k-distribution}
Log-log plots of momentum distributions found by QMC calculations in the strongly interacting regime.
(a)~Low temperature: $\xig=4.47$ and $\xiT=0.0085$. (b)~Temperature at the maximum contact: $\xig=1.26$ and $\xiT=0.49$.
The solid blue lines with shaded statistical error bars are the QMC results, the dashed red lines are algebraic fits to the large-$k$ tails, and the dotted green lines are the momentum distributions of the nondegenerate ideal gas. The insets show the same data in lin-lin scale.
}
 \end{figure}

Our predictions can be investigated with quantum gases where the Tan contact is extracted from radio-frequency spectra or momentum distributions~\cite{stewart2010,wild2012,luciuk2016,chang2016}.
Figure~\ref{fig:k-distribution} shows momentum distributions found from QMC calculations in the strongly interacting regime close to zero temperature [$\xiT \ll 1$, Fig.~\ref{fig:k-distribution}(a)] and at the contact maximum [$\xiT \simeq \xiT^*$, Fig.~\ref{fig:k-distribution}(b)].
In both cases, an algebraic decay at large momenta is observed,
with an amplitude matching our estimate for the contact~\cite{note:dk-fit}. 

In ultracold atom experiments, 1D systems are produced in the strongly interacting regime when loaded in 2D arrays of tubes~\cite{paredes2004, kinoshita2004,meinert2015},
which raises the question of the effect of averaging the momentum distributions over the tubes.
However, in the strongly interacting regime,
the relative amplitude of the maximum with respect to the zero-temperature value, $C^*/C^0$, is lower bounded by its value in the central tube~\cite{note:SM}.
For the parameters of Ref.~\cite{meinert2015}, we find $C^*/C^0 \gtrsim 5.1$, which should be sufficient for a clear identification of the maximum.
Moreover, the condition for the tubes to be in the quasi-1D regime, $\kB T \ll \hbar\omega_\perp$,
where $\omega_{\perp}$ is the transverse trap frequency, can be fulfilled using a strong-enough transverse confinement~\cite{note:SM}.


\lettersection{Conclusion}
Summarizing, we have provided a complete characterization of the Tan contact for the trapped Lieb-Liniger gas with arbitrary interaction strength, number of particles, temperature, and trap frequency. We have derived a universal scaling function of only two parameters and we have shown that it is in excellent agreement
with the numerically exact QMC results over a wide range of parameters. As a pivotal result, we found that the contact exhibits a
maximum versus the temperature for any interaction strength.
This behavior is mostly marked in the gas with large interactions and provides an unequivocal signature of the crossover to fermionization.
In outlook, the analysis of the Tan contact can be used to identify critical behaviors~\cite{chen2014}. It can further be extended to the excited states and multicomponent quantum systems~\cite{decamp2016a,decamp2016b,kormos2011,johnson2017,patu2017,patu2018}.

 \acknowledgments
P.V. acknowledges Mathias Albert for useful discussions.
This research was supported by the
European Commission FET-Proactive QUIC (H2020 Grant No.~641122) and Paris region DIM-SIRTEQ.
It was performed using HPC resources from  GENCI-CCRT/CINES (Grant No.~c2017056853).
Numerical calculations make use of the ALPS scheduler library and statistical analysis tools~\cite{troyer1998,ALPS2007,ALPS2011}. 
D.C. acknowledges support from the Institut Universitaire de France.


\begin{thebibliography}{10}
\bibitem{note:normC}
{Here, we use the normalization condition of the momentum distribution $\int
  \frac{dk}{2\pi} n(k)=N$.}
\bibitem{note:symbols}
{For Fig.~\ref{fig:contact_vs_N-T}(a), we used $\vert\aOneD\vert/\aHO = 9.5$
  (red squares), $0.032$ (red diamonds), $1.41$ (green squares), $0.14$ (green
  diamonds), $0.14$ (blue squares), and $2.02$ (blue diamonds). For
  Fig.~\ref{fig:contact_vs_N-T}(b), we used $\vert\aOneD\vert/\aHO=10$ (blue
  squares), $31.62$ (blue diamonds), $0.45$ (green squares), $1.41$ (green
  diamonds), $75.0$ (red squares) and $47.4$ (red diamonds).}
\bibitem{note:maximum-QD}
Quantum degeneracy effects tend to increase the value of $\nu$ for small values
  of $\xig$. In the asymptotic limit $\xig \rightarrow 0$, they become
  dominant. Then $G(\alpha) \simeq \pi^2/\sqrt{\vert\alpha\vert}$ and $\nu
  \simeq 1$. However, in this regime, the maximum is extremely weak and hardly
  visible in practice.
\bibitem{note:Maxwell}
{Equation~(\ref{eq:Maxwell}) follows from the sweep relation
  $C=({4m}/{\hbar^2}) \left.{\partial F}/{\partial \aOneD}\right\vert_{T,N}$
  and the thermodynamic definition of the entropy $S=-\left.\partial F/\partial
  T\right\vert_{\aOneD,N}$, where $F(N,T,\aOneD)=\Omega(\mu,T,\aOneD)+\mu N$ is
  the Helmoltz free energy.}
\bibitem{note:dk-fit}
Fitting the tails of the momentum distributions found from the QMC calculations
  by $n(k) \simeq C_\textrm{\tiny fit}/k^{p_\textrm{\tiny fit}}$, we obtain
  $p_\textrm{\tiny fit}$ and $\widetilde{C}_\textrm{\tiny fit} = \aHO^3
  C_\textrm{\tiny fit} / N^{5/2}$. For Fig.~\ref{fig:k-distribution}(a),
  $p_{\textrm{\tiny fit}}=3.80\pm0.20$ and $\widetilde{C}_{\textrm{\tiny
  fit}}=(1.01 \pm 0.14) \times 10^{-3}$ while $\widetilde{C} \simeq 0.97 \times
  10^{-3}$. For Fig.~\ref{fig:k-distribution}(b), $p_{\textrm{\tiny
  fit}}=3.72\pm0.10$ and $\widetilde{C}_{\mathrm{fit}}=0.23\pm0.04$ while
  $\widetilde{C} \simeq 0.22$. The agreement with the expected exponent $p=4$
  is better than $7\%$ and with the contact $C$ better than $5\%$.
\end{thebibliography}

  
 \renewcommand{\theequation}{S\arabic{equation}}
 \setcounter{equation}{0}
 \renewcommand{\thefigure}{S\arabic{figure}}
 \setcounter{figure}{0}
 \renewcommand{\thesection}{S\arabic{section}}
 \setcounter{section}{0}
 \onecolumngrid  
     
 
 \newpage

 {\center \bf \large Supplemental Material for \\}
 {\center \bf \large Tan's Contact for Trapped Lieb-Liniger Bosons at Finite Temperature \\ \vspace*{1.cm}
 }

In this supplemental material, we provide details about the QMC approach (Sec.~\ref{sec:SMQMC}),
the derivation of the virial-expansion formula (Sec.~\ref{sec:SMQvirial}),
and a detailed analysis of experimental observability of the contact maximum in experiments using a 2D array of 1D tubes in strongly-interacting regime (Sec.~\ref{sec:1Dtube}).

\section{Quantum Monte Carlo simulations}
\label{sec:SMQMC}

\subsection{Path-integral Monte-Carlo approach}

The quantum Monte Carlo (QMC) calculations exploit the same implementation as detailed in Refs.~\cite{carleo2013,boeris2016}. The continuous-space path integral formulation allows us to simulate the exact Hamiltonian, Eq.~(1) of the main paper, for an arbitrary trap $V(x)$, within the grand-canonical ensemble. The statistical average of the number of world lines yields the total number of particles $N$, and the interaction energy $\langle \mathcal{H}_{\mathrm{int}}\rangle$ is computed from the zero-range two-body correlator. The contact is then found using the thermodynamic relation
\begin{equation}\label{eq:HintC}
C= ({2gm^2}/{\hbar^4})\langle \mathcal{H}_{\mathrm{int}}\rangle.
\end{equation}
The world lines are discretized into an adjustable number $M$ of slices of elementary imaginary propagation time $\epsilon=1/M \kB T$ each, and sampled using the worm algorithm~\cite{boninsegni2006,Boninsegni2006b}. Each calculation is run for various values of $\epsilon$ and polynomial extrapolation is used to eliminate systematic finite-time discretization errors (see below).

\subsection{Finite-$\boldsymbol{\epsilon}$ scaling}

\begin{figure}[b!]
\vspace{-0.5cm}
\centering
\includegraphics[width=0.45\textwidth]{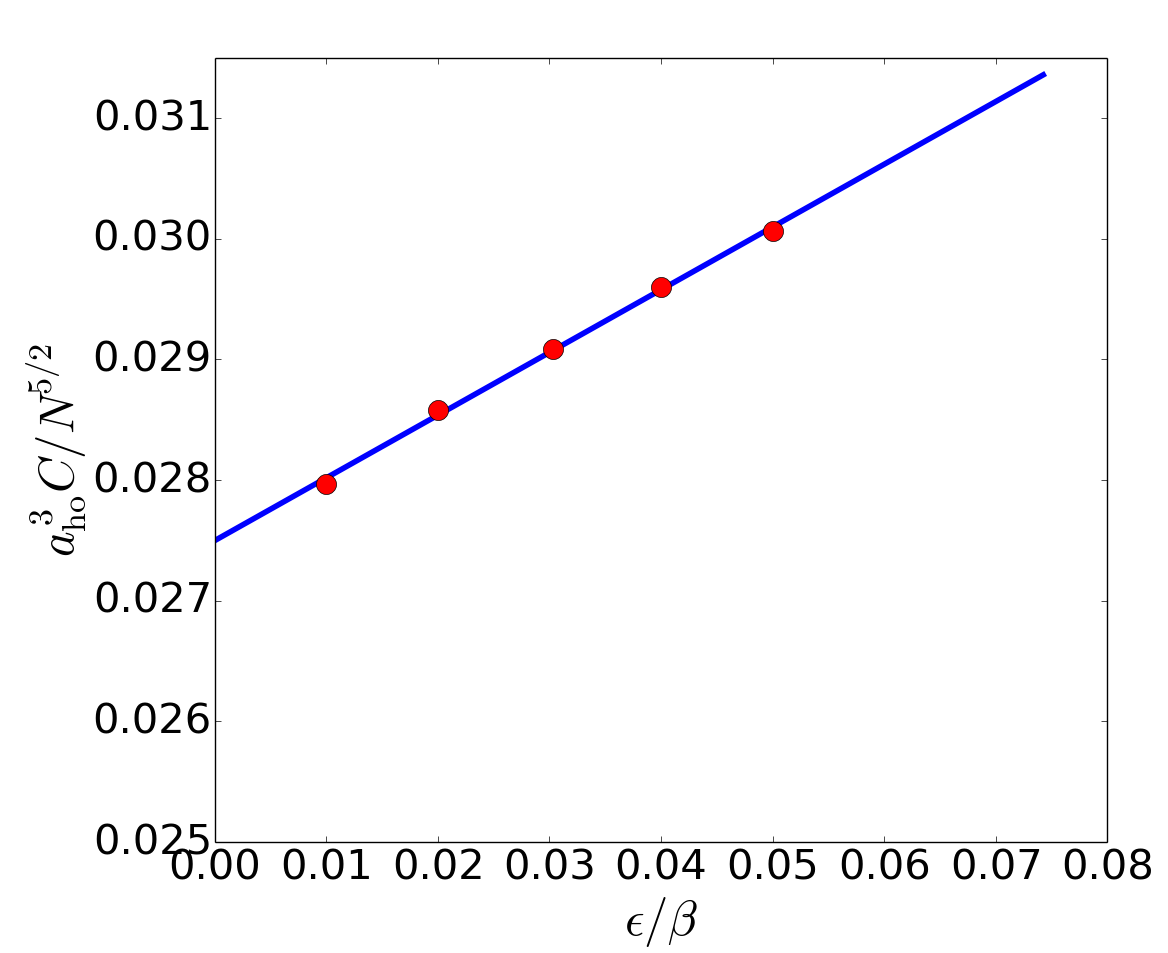}
\includegraphics[width=0.445\textwidth]{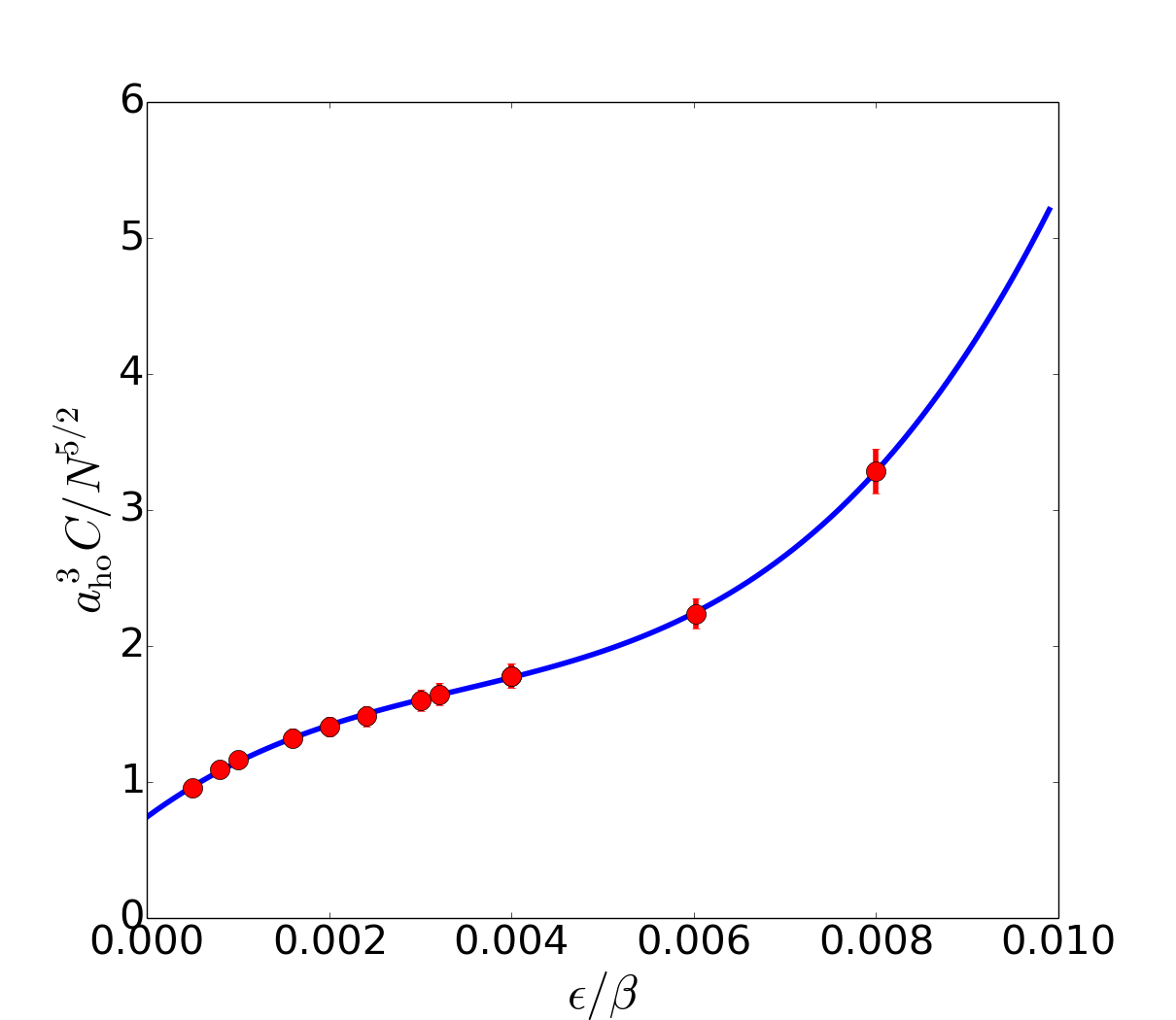}
\caption{
Quantum Monte Carlo (QMC) results for the reduced Tan contact
for $\xiT=|\aOneD|/\lambdaT = 0.28$ and $\xig=\aHO/\vert\aOneD\vert\sqrt{N}=0.1$ (left panel)
and for $\xiT = 0.0085$ and $\xig=4.47$ (right panel).
The red points show the QMC results for various values of the dimensionless parameter $\epsilon/\beta$, where $\beta=1/\kB T$ is the inverse temperature,
togther with a linear (left panel) or third-order polynomial (right panel) fit.
}
\label{Fig.lable}
\end{figure}

The QMC results are exact in the $\epsilon\rightarrow 0$ limit.
In order to find the final results reported on the Fig.~2 of the main paper, we proceed as follows.
For each set of physical parameters (interaction strength, chemical potential, temperature, and trap frequency), we perform a series of QMC calculations for different values of $\epsilon$ and extrapolate the result to the limit $\epsilon\rightarrow 0$.

For most of the calculations, we are able to use a sufficiently small value of $\epsilon$ and a linear extrapolation is sufficient. We fit the QMC data with $\aHO^3C/N^{5/2}=a + b(\epsilon/\beta)$, with $a$ and $b$ as fitting parameters. We then use the quantity $a$ as the final result for $\aHO^3C/N^{5/2}$. 
An example is shown on the left panel of Fig.~\ref{Fig.lable} below. In this case, the linear extrapolation only corrects the QMC result for the smallest value of $\epsilon\ (\epsilon/\beta = 0.01)$ by less than $4\%$.

In some cases, however, the linear fit is not sufficient for extrapolating correctly the QMC results. This occurs in the strongly-interacting regime for low to intermediate temperatures. In such cases, we use a third-order polynomial, $\aHO^3C/N^{5/2}=a + b(\epsilon/\beta) + c(\epsilon/\beta)^2+d(\epsilon/\beta)^3$, to extrapolate the finite-$\epsilon$ numerical data.
An example is shown on the right panel of Fig.~\ref{Fig.lable}. In this case, the extrapolation corrects the QMC result for the smallest value of $\epsilon\ (\epsilon/\beta = 0.0005)$ by roughly $25\%$.

For all the QMC results reported on the Fig.~2 of the main paper, we have performed a systematic third-order polynomial extrapolation, even when a linear extrapolation was sufficient.

\section{Derivation of the Tan contact  in the large-temperature and large-interaction limit}
\label{sec:SMQvirial}

We derive here Eq.~(8) of the main paper for the contact at large, finite temperature ($\kB T\gg N \hbar\omega$) and large interactions $|\aOneD|/\lHO\ll1$. As a first step, the Tan contact at large temperature and arbitrary interactions can be estimated using the first term of the virial expansion~\cite{vignolo2013}
\begin{equation}
{C}=\dfrac{4m\omega}{\hbar \lambdaT}N^2\,c_2
\label{eq:c}
\end{equation}
where $c_2=\lambdaT\dfrac{\partial b_2}{\partial |\aOneD|}$ and
$b_2=\sum_\nu e^{-\beta\hbar\omega(\nu+1/2)}$.
The $\nu$'s are the solutions of the transcendental equation
\begin{equation}
f(\nu)=\dfrac{\Gamma(-\nu/2)}{\Gamma(-\nu/2+1/2)}=\sqrt{2}\dfrac{\aOneD}{\lHO}.
\label{bush}
\end{equation}
By exploiting the Euler reflection formula
\begin{equation}
\Gamma(z)\Gamma(1-z)=\dfrac{\pi}{\sin(\pi z)},
\label{gammarel}
\end{equation}
one can re-write Eq. (\ref{bush}) under the form
\begin{equation}
f(\nu)=-{\rm cot}(\pi\nu/2)\dfrac{\Gamma(\nu/2+1/2)}{\Gamma(\nu/2+1)}.
\end{equation}
By using the asympotic expansions
\begin{equation}
  \Gamma(z)\simeq\sqrt{2\pi}\, z^{z-1/2}\, e^{-z}\left(1+\dfrac{1}{12\, z}+O(1/z^2)\right)
\label{exp1}
\end{equation}
and
\begin{equation}
  \Gamma(z+1/2)\simeq\sqrt{2\pi}\, z^{z}\, e^{-z} \left(1-\dfrac{1}{24\, z}+O(1/z^2)\right),
\label{exp2}
\end{equation}
we obtain the following asymptotic expression for $f(\nu)$
\begin{equation}
   f(\nu) \simeq -{\rm cot}(\pi\nu/2) \frac{1}{\sqrt{\nu/2+1/2}}
    \simeq -\sqrt{\frac{2}{\nu}} {\rm cot}(\pi\nu/2).
\end{equation}
In the Tonks-Girardeau regime, corresponding to  $\aOneD=0$, one has $\nu=2n+1$, with $n\in\mathbb{N}$. Thus, in the regime $|\aOneD|/\lHO\ll1$, we obtain an explicit expression for $\nu$, by writing
\begin{equation}
\sqrt{\dfrac{2}{2n+1}}\cot(\pi\nu/2)\simeq\sqrt{2}\dfrac{\vert\aOneD\vert}{\lHO}
\end{equation}
namely
\begin{equation}
\nu_n = \dfrac{2}{\pi}{\rm{acot}}(\sqrt{2n+1}|\aOneD|/\lHO)+2n, \qquad n\in\mathbb{N}.
\end{equation}
This yields the following explicit expression for $c_2$:
\begin{equation}
  \begin{split}
    c_2&=\lambdaT\sum_\nu(-\beta\hbar\omega)\dfrac{\partial\nu}{\partial|\aOneD|}e^{-\beta\hbar\omega(\nu_n+1/2)}\\
      &=\lambdaT\sum_n(-\beta\hbar\omega)\dfrac{2}{\pi}\dfrac{\sqrt{2n+1}}{\lHO}
    \dfrac{-1}{1+(2n+1)\frac{\aOneD^2}{\lHO^2}}e^{-\beta\hbar\omega(\nu_n+1/2)}\\
    &=\dfrac{2\lambdaT\beta\hbar\omega}{\pi\lHO}\sum_n\dfrac{\sqrt{2n+1}}{1+(2n+1)\frac{\aOneD^2}{\lHO^2}}e^{-\beta\hbar\omega(\nu_n+1/2)}.
\label{sumc2}
  \end{split}
\end{equation}
In order to evaluate analytically the sum in Eq. (\ref{sumc2}), we replace $\nu$ with $2n+1$ in the
exponential. Indeed, the first-order correction in
$|\aOneD|$
  gives a negligible contribution in the limit $\beta\rightarrow 0$ and
  $\aOneD\rightarrow 0$.
  We finally get 
  \begin{equation}
  \begin{split}
  c_2&=\sqrt{2}\left(\dfrac{1}{2\pi\xiT^2}-\dfrac{e^{1/2\pi\xiT^2}}{2^{3/2}\pi\xiT^3}{\rm Erfc}(1/\sqrt{2\pi}\xiT)\right).\\
\end{split}
  \end{equation}
  
  \begin{figure}
  \begin{center}
    \includegraphics[width=0.7\linewidth]{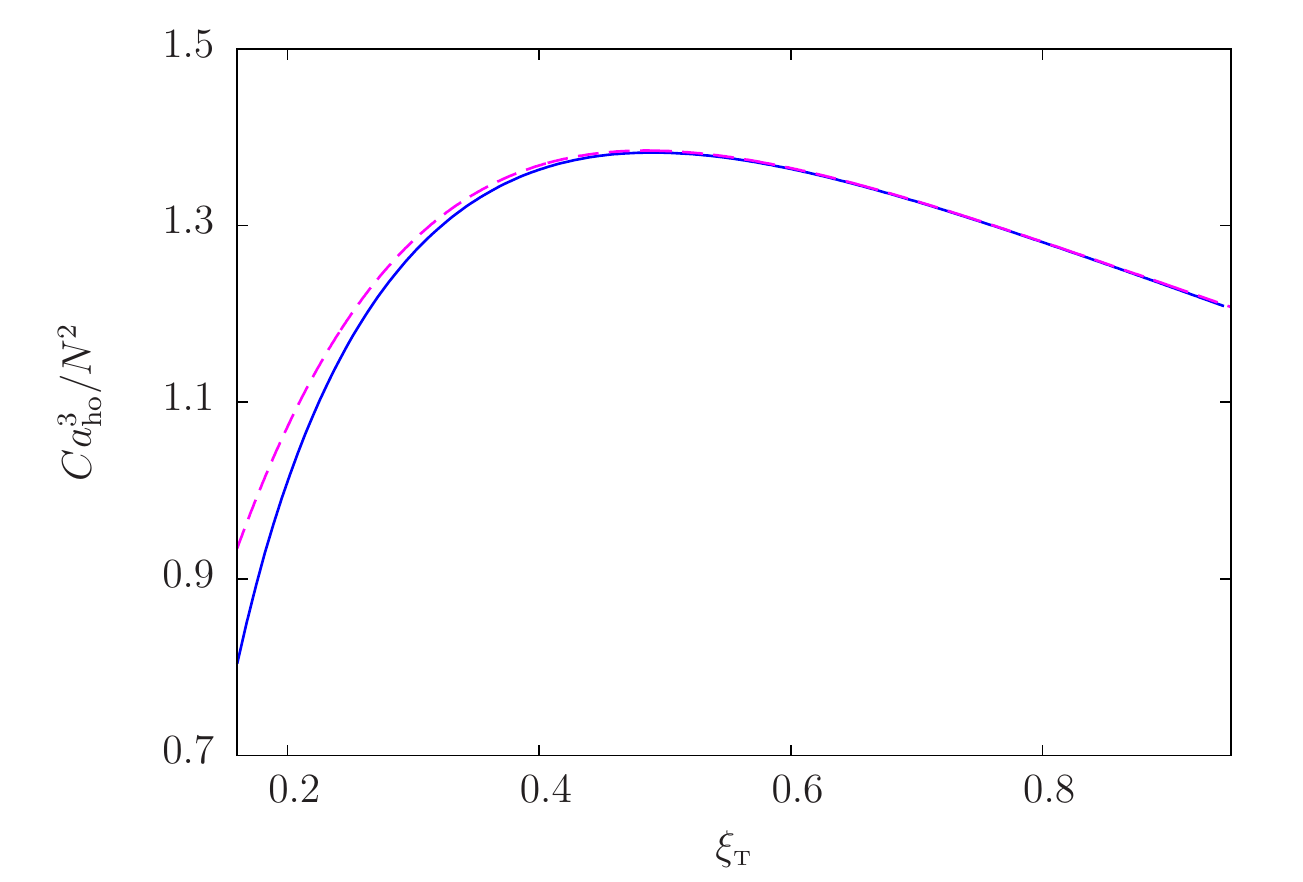}
    \end{center}
\caption{\label{figsuppl1} $C\lHO^3/N^2$ calculated using the numerical solution of Eq.~(\ref{bush}) (blue continuous line) and using the analytical expression (\ref{formula-pat}) (magenta dashed line). We have considered $|\aOneD|/\aHO=0.4$.}
\end{figure}

  Thus, the contact at large temperatures and large interactions can be approximated by
    \begin{equation}
      \begin{split}
        C=&\dfrac{4\sqrt{2}N^2\xiT}{|\aOneD|\lHO^2}\left(\dfrac{1}{2\pi\xiT^2}-\dfrac{1}{2^{3/2}\pi\xiT^3}e^{1/2\pi\xiT^2}{\rm Erfc}(1/\sqrt{2\pi}\xiT)\right)\\
        =&\dfrac{2N^{5/2}}{\pi\lHO^3}\dfrac{\xig}{\xiT}\left(\sqrt{2}-\dfrac{\mathrm{e}^{1/2\pi\xiT^2}}{\xiT}{\rm Erfc}(1/\sqrt{2\pi}\xiT)\right).
    \label{formula-pat}
\end{split}
      \end{equation}
We have checked that this expression is in excellent agreement with the calculation of Eq.~(\ref{eq:c}) using the numerical solution of Eq.~(\ref{bush}), see  Fig.~\ref{figsuppl1}.

\section{Experimental observability of the maximum contact in a 2D array of 1D tubes}
\label{sec:1Dtube}

We address here a more detailed explanation for the question of observability of the contact maximum in ultracold-atom experiments. In subsection~\ref{sec:averaging}, we discuss the effect of averaging the momentum distribution over the 2D array of tubes. In subsection~\ref{sec:quasi1D}, we discuss the validity of the quasi-1D gas condition.

\subsection{Averaging over the tubes}
\label{sec:averaging}

In most cases, strong transverse confinement is realized by applying a 2D optical lattice in the directions $y$ and $z$, orthogonal to the 1D direction $x$.
For sufficiently strong lattices, it creates an array of independent 1D tubes, indexed by the labels $(j,\ell)\in\mathbb{Z}^2$.
The total contact then reads as
 \begin{equation}
C = \sum_{j,\ell} C(j,\ell),
\end{equation}
where $C(j,\ell)$ is the contact in the corresponding tube.
Each tube is populated with a number $N(j,\ell)$ of atoms, which depends on the loading procedure of the atoms in the 2D lattice.
Since the number of atoms is maximum in the central tube ($j=\ell=0$), we have $\xig(j,\ell) \geq \xig(0,0)$,
and the condition for having all tubes in the strongly-interacting regime reduces to $\xig(0,0) \gg 1$.

In that regime, the temperature dependence of the contact around the maximum is independent of $\xig$, and thus independent of the tube.
Indeed, as shown by Eq.~(8) of the main paper, the parameter $\xig$ just appears as a prefactor.
In particular, the maximum contact is located at the universal value $\xiT^* \simeq 0.485$, which is identical for the tubes.
Using Eq.~(8) of the main paper, we then find
 \begin{equation}\label{eq:Cstar}
C^* \simeq 0.55 \times \sum_{j,\ell} \frac{N(j,\ell)^{5/2}\xig(j,\ell)}{\aHO^3}.
\end{equation}
At zero temperature, the contat may be found using the mapping between the strongly-interacting Bose gas and the strongly-degenerate ideal Fermi gas.
It yields the value~\cite{olshanii2003}
 \begin{equation}\label{eq:Czero}
C^0 \simeq 0.82 \times \sum_{j,\ell} \frac{N(j,\ell)^{5/2}}{\aHO^3}.
\end{equation}
We then find that the relative amplitude of the maximum contact with respect to its zero-temperature value fullfils the inequality
\begin{equation}\label{eq:LowerBound}
\frac{C^*}{C^0} \gtrsim 0.68 \times \xig(0,0).
\end{equation}
Therefore, the relative ampitude of the maximum contact is larger than a fraction of the interaction parameter $\xig(0,0) \gg 1$ and should be observable.
For instance, for the parameters of Ref.~\cite{meinert2015}, we find $\xig(0,0) \simeq 7.5$ and $C^*/C^0 \gtrsim 5.1$.

Note that the lower bound in Eq.~(\ref{eq:LowerBound}) is universal in the sense that it does not depend on the distribution of atoms in the various lattice tubes.
Note also that it is immune to shot-to-shot fluctuations of the atom numbers in the tubes.

Finally, a more precise value of the relative amplitude of the maximum contact is found by compting the sums in Eqs.~(\ref{eq:Cstar}) and (\ref{eq:Czero}) for realistic distributions of the atom numbers among the tubes. Using the estimate
 \begin{equation}
N_{j,\ell} = \left[1-\frac{2\pi N(0,0)}{5N}\big(j^2+\ell^2\big)\right]^{3/2},
\end{equation}
relevant to the experiments of Refs.~\cite{paredes2004,meinert2015}, we find $C^*/C^0 \simeq 0.8\times\xig(0,0)$, which is only about $20\%$ larger than the atom distribution-independent lower bound, Eq.~(\ref{eq:LowerBound}). For the parameters of Ref.~\cite{meinert2015}, it yields $C^*/C^0 \gtrsim 6.1$.  It may be further increased by lowering the total number of atoms, although at the expense of atom detectability.
In this respect, mestable He atoms appear particularly suited for they allow for atom-resolved detection and measurement of momentum distributions over up to six decades~\cite{chang2016}. 

\subsection{Validity condition of the quasi-1D regime}
\label{sec:quasi1D}
  
\begin{figure}
\begin{center}
\includegraphics[width=0.5\linewidth]{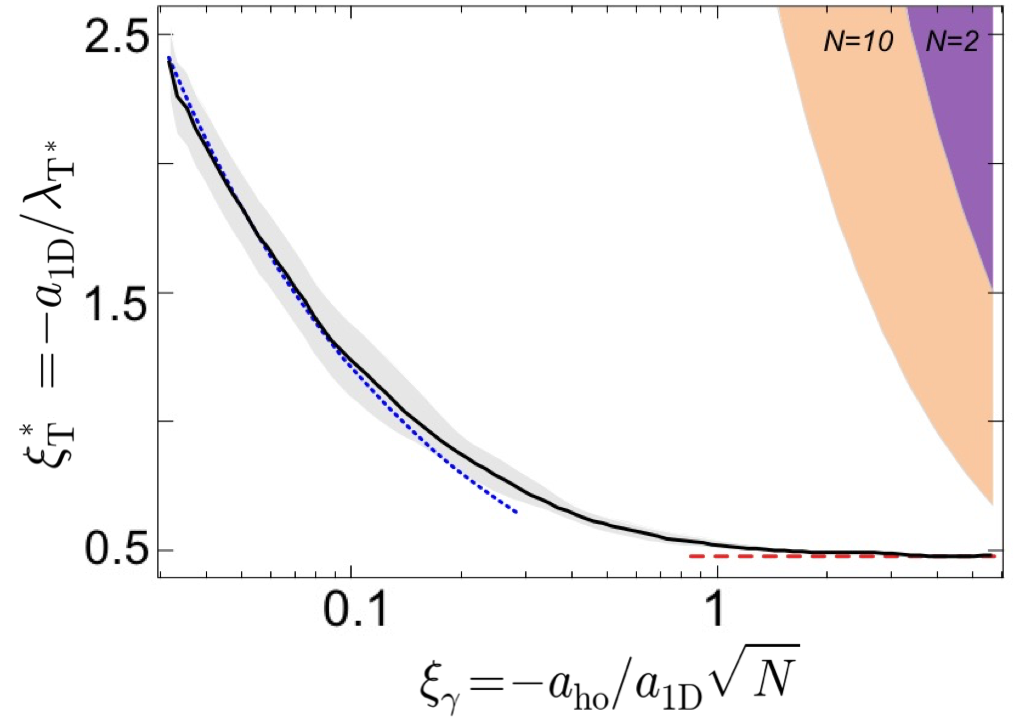}
\end{center}
\caption{\label{figsuppl1}
Reproduction of Fig.~3 of the main paper together with the validity condition of the quasi-1D regime, Eq.~(\ref{eq:1Dcondition}), for $\sqrt{\omega_{\perp}/\omega} = 30$, relevant for the experiments of Ref.~\cite{meinert2015}.
The dark regions show the excluded regions for $N=2$ atoms (black) and $N=10$ atoms (gray).
}
\end{figure}

The condition for generating truly quasi-1D tubes in the experiment reads as $\kB T\ll \hbar\omega_{\perp}$, where $\omega_{\perp}$ is in the angular frequency of the transverse confinement induced by the 2D lattice, see for instance Ref.~\cite{olshanii1998}.
This condition may be written using the scaling parameters $\xig=-\aHO/\aOneD\sqrt{N}$ and $\xiT=-\aOneD/\lambdadB$. Using the relations $\aHO=\sqrt{\hbar/m\omega}$, $\lambdaT = \sqrt{2\pi\hbar^2/m\kB T}$, and
\begin{equation}
\xiT\xig=\frac{1}{\sqrt{2\pi N}}\sqrt{\frac{\kB T}{\hbar\omega}},
\end{equation}
it reads as
\begin{equation}\label{eq:1Dcondition}
\xiT\xig\ll \frac{1}{\sqrt{2\pi N}}\times \sqrt{\frac{\omega_{\perp}}{\omega}}.
\end{equation}
In experiments, the typical value of $\omega_{\perp}/\omega$ varies from a few hundreds to a few thousands.
In Fig.~\ref{figsuppl1} we reproduce the Fig.~3 of the main paper, together with the condition~(\ref{eq:1Dcondition}) for two values of the atom number $N$ and the parameters of Ref.~\cite{meinert2015}, $\omega/2\pi = 15.8$Hz and $\omega_\perp/2\pi = 14.5$kHz. The regions where the validity condition is not fulfilled is shown in black for $N=2$ and gray for $N=10$. We conclude that the value $\xiT^*$ corresponding to the maximum of the contact is well inside the validity regime deep enough in the strongly-interacting regime, $\xig \gg 1$. It is thus possible to observe the maximum contact in this regime. Moreover, one can further extend the validity region by increasing the value of the ratio $\omega_{\perp}/\omega$, \ie\ by increasing the transverse confinement.

\end{document}